# Positive exchange-bias in ferromagnetic $La_{0.67}Sr_{0.33}MnO_3$ / $SrRuO_3$ bilayers

X. Ke and M.S. Rzchowski

*Physics Department, University of Wisconsin-Madison,*

*Madison, WI 53706, USA*

L.J. Belenky and C.B. Eom

*Department of Materials Science and Engineering,*

*University of Wisconsin-Madison, Madison, WI 53706, USA*

Epitaxial $La_{0.67}Sr_{0.33}MnO_3$ (LSMO)/ $SrRuO_3$ (SRO) ferromagnetic bilayers have been grown on (001) $SrTiO_3$ (STO) substrates by pulsed laser deposition with atomic layer control. We observe a shift in the magnetic hysteresis loop of the LSMO layer in the same direction as the applied biasing field (positive exchange bias). The effect is not present above the Curie temperature of the SRO layer ($T_c^{SRO}$), and its magnitude increases rapidly as the temperature is lowered below $T_c^{SRO}$. The direction of the shift is consistent with an antiferromagnetic exchange coupling between the ferromagnetic LSMO layer and the ferromagnetic SRO layer. We propose that atomic layer charge transfer modifies the electronic state at the interface, resulting in the observed antiferromagnetic interfacial exchange coupling.

Interactions at magnetic interfaces are central to the operation of virtually all magnetic heterostructures. When the interface is between two magnetic materials, the exchange interaction between spins at the interface is dominant, and can dramatically change the magnetic response of the overall heterostructure. Interfacial interactions in bilayers of two ferromagnets can couple the layers so strongly that both switch as a single unit[1]. In a bilayer of two ferromagnets with very different coercive fields, an 'exchange spring' effect[2] can result, where the magnetization of the soft layer can reversibly 'twist' with respect to the hard layer. The interfacial interaction can also shift the magnetic hysteresis loop so that it is no longer symmetric about zero applied field. This exchange bias effect, most commonly implemented by a ferromagnet/antiferromagnet interface, has been extensively used to pin the magnetization in one of the layers of a magnetic spin-valve or tunnel junction.

Magnetization loop shifts of a ferromagnetic film arising from interfacial exchange interactions with a 'pinning layer' have been observed in many systems, including ones in which the pinning layer is ferromagnetic[3] or ferrimagnetic[4], as well as antiferromagnetic[5]. Although the details of the interfacial spin arrangements are in some cases believed to be quite complicated, in all cases a preferred direction of the pinning layer is induced by the application of a 'bias' magnetic field. In the case of an antiferromagnetic (AF) pinning layer, the dipole interaction with the applied field is usually small, and the preferred direction is determined by interaction of the AF layer with the ferromagnetic layer. An applied field can directly manipulate the magnetic orientation of ferro- and ferrimagnetic pinning layers through the dipole interaction.

These interactions are characterized by the exchange field $H_E$, the shift from zero of the magnetization loop center along the field axis. In almost all cases, this shift of the magnetization loop is opposite to the direction of applied bias field. This case is sometimes referred to as negative exchange bias. This amount of shift is the exchange field $H_E$. Recently there have been reports of *positive* exchange bias, in which the interfacial exchange interaction is believed to be antiferromagnetic. The origin of such antiferromagnetic coupling as proposed has been system-specific, but related to a microscopic antiferromagnetic coupling already present within the pinning layer.

Most studies of exchange bias have been done with bilayers of metallic ferromagnets and transition-metal antiferromagnetic oxides or alloys[5]. Perovskite magnetic oxide materials offer several advantages for exchange bias studies. The magnetic properties can be changed by cation substitution, similar lattice constants in magnetically different materials permit epitaxial growth, and interfaces between dissimilar materials can be made atomically abrupt. Negative exchange bias in epitaxial multilayers comprised of ferromagnet $La_{2/3}Ca_{1/3}MnO_3$ and antiferromagnet $La_{1/3}Ca_{2/3}MnO_3$ has recently been reported[6], as well as in $SrRuO_3/Sr_2YRuO_6$ ferro/antiferro bilayers[7].

Here we report the observation of *positive* exchange bias in epitaxial $La_{2/3}Sr_{1/3}MnO_3$ (LSMO) / $SrRuO_3$ (SRO) bilayers. Both layers are ferromagnetic as isolated films, with $T_c^{SRO} < T_c^{LSMO}$. The sense of the exchange bias we observe is always positive, and the effect is observed only below the Curie temperature of the SRO layer. We observe this effect independent of the order of the layers in bilayers films. We argue that antiferromagnetic interfacial coupling is the origin of the positive exchange bias. We attribute it to a change in the interfacial charge distribution arising from atomic layer charge transfer, and the strong dependence of magnetic properties on hole doping demonstrated in LSMO[8].

LSMO is a member of the colossal magnetoresistance (CMR)[9] perovskite oxide family $La_{1-x}R_xMnO_3$ (R=Sr, Ca, Ba, Pr). It has a relatively high curie temperature (bulk $T_c$~350K), small magnetic anisotropy and low coercive field[10]. SRO is also a perovskite ferromagnetic oxide, but one with a strong uniaxial crystalline anisotropy and relatively large (~1 T) coercive field[11], a relatively small moment (~1.5$\mu_B$ per Ru)[12], and a low Curie temperature (bulk $T_c$~150K). These factors permit switching the LSMO magnetization by sweeping a small field while keeping the SRO layer pinned. Both materials have metallic conductivity in the temperature range of our measurements.

Epitaxial bilayers were grown on (001) SrTiO$_3$ substrates by pulsed laser deposition (PLD) using a KrF excimer laser (248 nm) in 100 mTorr of oxygen with *in-situ* high pressure reflection high-energy electron diffraction (RHEED) as described previously[13]. The LSMO layers were grown at 850ºC and the SRO layers at 600ºC for bilayers samples. After deposition, samples were cooled to room temperature in a 300 Torr O$_2$ environment. During growth, the thickness and growth mode were monitored with both RHEED intensity modulation and diffraction pattern. Clear oscillations of the specular spot were observed during growth of both materials (see Fig. 1), permitting precise control of the layer thicknesses. Two types of bilayer samples were measured. Both consisted of one layer of LSMO (thickness 200 Å) and one layer of SRO (thickness 100 Å). In the first sample type, the SRO layer was first grown on the (001) STO substrate, followed by growth of the LSMO layer on top of the SRO. In the second sample type, the order of the layers was reversed, with SRO being the top of the bilayer. All layers were found to be epitaxial. Rocking curves on all layers had FWHM less than 0.17˚, indicating excellent crystallinity.

*Ex-situ* AFM measurements showed a terrace surface morphology to the topmost layer in both types of bilayers, with unit cell terrace heights and terrace widths consistent with the 0.2˚ miscut of the SrTiO$_3$ substrate. X–ray diffraction measurements show clear epitaxial orientation of the layers with respect to the substrate, and some strain relaxation in the layers. The bulk pseudocubic lattice constants are 3.87 Å (LSMO), 3.93 Å (SRO), and 3.905 Å (STO). Thus the SRO layers are expected to be under compressive strain, and the LSMO layers under tensile strain. This corresponds to out-of-plane lattice constants reduced from the bulk value for LSMO, and greater than the bulk value for SRO. The out-of-plane lattice constants of the LSMO top-layer sample were determined to be 3.845 Å (LSMO) and 3.945 Å (SRO). For the SRO top-layer sample they were determined to be 3.852 Å (LSMO) and 3.957 Å (SRO). In both samples this indicates greater strain in a layer when it is topmost than when it is grown directly on the substrate. This is consistent with partial strain relaxation in the bottom layer, leading to a growth template with a larger lattice mismatch than the STO substrate. In all samples the measured Curie temperature of LSMO is ~350 K, while that of SRO is ~110 K. The reason for suppressed $T_c$ of SRO is not completely understood, but the known volatility of Ru may have resulted in some nonstochiometry.

Exchange bias properties of the samples were studied in the temperature range of 5–400 K by a Quantum Design superconducting quantum interference device (SQUID) magnetometer. Samples were first cooled in zero field to 10 K, after which the SrRuO$_3$ layer magnetization was pinned by application of a several Tesla magnetic field. This biasing technique is generally not possible for an antiferromagnetic pinning layer, which does not couple directly to the applied field. Hysteresis loops were then measured by sweeping the applied field over the range ±1500 Oe. These technically are minor loops of the bilayers, but full loops of the LSMO layer, as the maximum field is far below that required to reverse the SRO layer. Figure 2 shows data at 10 K for the bilayer sample with SRO top layer, and Fig. 3 shows data for the bilayer with LSMO top layer. In both figures, panel (*a*) shows the loop after application of –3 T field (and subsequent ramp to –1500 Oe), and panel (*b*) after application of +4 T field (and subsequent ramp to +1500 Oe).

In all cases the magnetization loop is shifted along the field axis in the same direction as the applied magnetic field. The shifted hysteresis loop provides direct evidence of positive exchange bias in LSMO/SRO bilayers. The LSMO layer dominates the magnetization signal, and so little vertical shift from the SRO is seen. We have also measured 10 K magnetization loops after cooling through $T_c^{SRO}$ in various fields. We find the exchange field $H_E$ largely independent of cooling field, with $H_E$ beginning to change sign near zero cooling field.

Figure 4 shows the temperature dependence of the exchange field $H_E$ for SRO/LSMO//STO bilayer after pinning the SRO layer with –3 Telsa. The absolute value of the exchange field decreases quickly with increasing temperature and is zero within the 10 Oe measurement resolution for $T \gtrsim T_c^{SRO}$. This is very similar to the behavior of exchange bias at a ferromagnetic/antiferromagnetic material interface. Both types of bilayers show nominally the same exchange field. The only clear differences are a wider and less sharp hysteresis loop of the LSMO top-layer bilayer relative to the SRO top-layer bilayer (see Figs 2 and 3). That is, the coercivity of LSMO is larger when LSMO is grown on SRO than when grown on the STO substrate. We attribute this to the larger lattice mismatch between LSMO and SRO and potential partial relaxation.

We have verified that the coupling between the layers is not magnetostatic by growing a trilayer magnetic tunnel junction configuration SRO(400 Å)/LSAT(20 Å)/LSMO(400 Å)//STO, where LSAT is the nonmagnetic insulator $(LaAlO_3)_{0.3}(Sr_2AlTaO_6)_{0.7}$. The growth conditions were the same as for the bilayers. The magnetization loop of this trilayer shows no shift at 10 K, which means that the LSMO layer is not biased. This demonstrates that magnetostatic couplings are small, and suggests that the loop shift arises from the LSMO / SRO interface.

The relation between antiferromagnetic interfacial coupling and positive exchange bias is apparent in a simple model where reversal of the LSMO layer is through rigid, coherent rotation. A large enough positive field will align the magnetizations of both layers in the positive direction through the dipole interaction $\vec{M} \bullet \vec{H}$. As the applied field is decreased, the SRO magnetization remains fixed due to the strong anisotropy, but the LSMO layer can rotate. Without an exchange interaction at the interface ($J_{ex}$=0), reversal of the LSMO layer occurs at $H$=0. An antiferromagnetic interfacial exchange interaction ($J_{ex}$<0) favors antiparallel alignment, and reversal occurs earlier in the loop, at a positive field $H = -J_{ex}/M$. Completely rigid rotation is likely not the case in LSMO, but rather the layer will incorporate partial or full domain walls as the magnetization twists[4]. A magnetocrystalline anisotropy can alter the magnitude of the exchange bias[14]. Domain wall pinning can lead to an intrinsic coercivity, altered by exchange bias[15]. These effects have been shown to modify aspects of the magnetization reversal process, but the broad correlation between antiferromagnetic coupling and positive exchange bias remains[5].

An interesting aspect is the origin of an antiferromagnetic coupling between spins at the interface when the materials themselves have ferromagnetic interactions. In ferrimagnetic bilayers, positive exchange has been attributed to a microscopic antiferromagnetic coupling between the two (unequal) components of the magnetization, consistent with the interactions within the layers[4]. In Fe(ferro)/FeF$_2$(antiferro) bilayers, the antiferromagnetic interfacial coupling has been proposed to arise through the same fluorine superexchange as in the antiferromagnetic layer[16]. In our LSMO/SRO bilayers, the microscopic exchange is *ferromagnetic* within both layers.

However the interface is likely to have a different electronic and magnetic structure, due to altered interfacial bond geometry and to altered interfacial charge transfer. Mn and Ru ions are separated at the interface by either a SrO or (LaSr)O layer. The Mn-O-Ru bond angle through this layer will be different than in either film. To the extent that superexchange influences the interfacial exchange, the interfacial exchange will be different from that in either film. The interfacial electronic state will also be different. The $SrRuO_3$ film is composed of individually charge-neutral $Sr^{2+}O^{2-}$ and $Ru^{4+}(O^{2-})_2$ atomic layers. The atomic layers in $LaSrMnO_3$ [$(La_{1-x}Sr_x)^{(3-x)+}O^{2-}$ and $Mn^{3+}(O^{2-})_2$] are formally charged, attaining neutrality through interlayer charge transfer. At the $MnO_2$/SrO or (LaSr)O/$RuO_2$ interface between the films, the different valence balance leads to different charge transfer and potentially a different local hole concentration. The sensitivity of the effective LSMO exchange coupling to hole doping[8] suggests that the interfacial coupling could be different from the bulk. Depending on the range of the charge transfer, this could be a purely interfacial antiferromagnetic exchange, or a several unit cell antiferromagnetic interfacial layer. Similar charge transfer has been suggested as the origin of unexpected magnetic configurations at other perovskite oxide interfaces[17].

This work was supported by the National Science Foundation through the NIRT program, award No. 0210449.

FIGURE CAPTIONS

Figure 1: RHEED intensity during epitaxial growth. Strong oscillations during LSMO growth indicate a layer-by-layer growth mode. Intensity recovery and rapid oscillations during SRO growth are consistent with step-flow growth. Diffraction patterns with Kikuchi lines before growth (a) after LSMO (b) and after SRO (c) indicate epitaxial registry.

Figure 2: Hysteresis loop at 10 K for SRO/LSMO bilayer grown on STO substrate with SRO on the top (SRO/LSMO//STO). The magnetization loop is shifted to negative field direction with –3 Telsa field applied along the surface before ramping the field (a); with +4 Telsa field applied, loops are shifted to positive field direction (b).

Figure 3: Hysteresis loop at 10 K for LSMO/SRO bilayer grown on STO substrate with LSMO on the top (LSMO/SRO//STO). The magnetization loop is shifted to negative field direction with –3 Telsa filed applied along the surface before ramping the field (a); with +4 Telsa field applied, loops are shifted to positive field direction (b).

Figure 4: Exchange field *vs* temperature for SRO/LSMO//STO bilayer with -3 Telsa field applied along the surface before ramping the field. The exchange field decreases with increasing temperature and disappears above $T_c^{SRO}$. There is ~10 Oe residual due to a coarse field step.

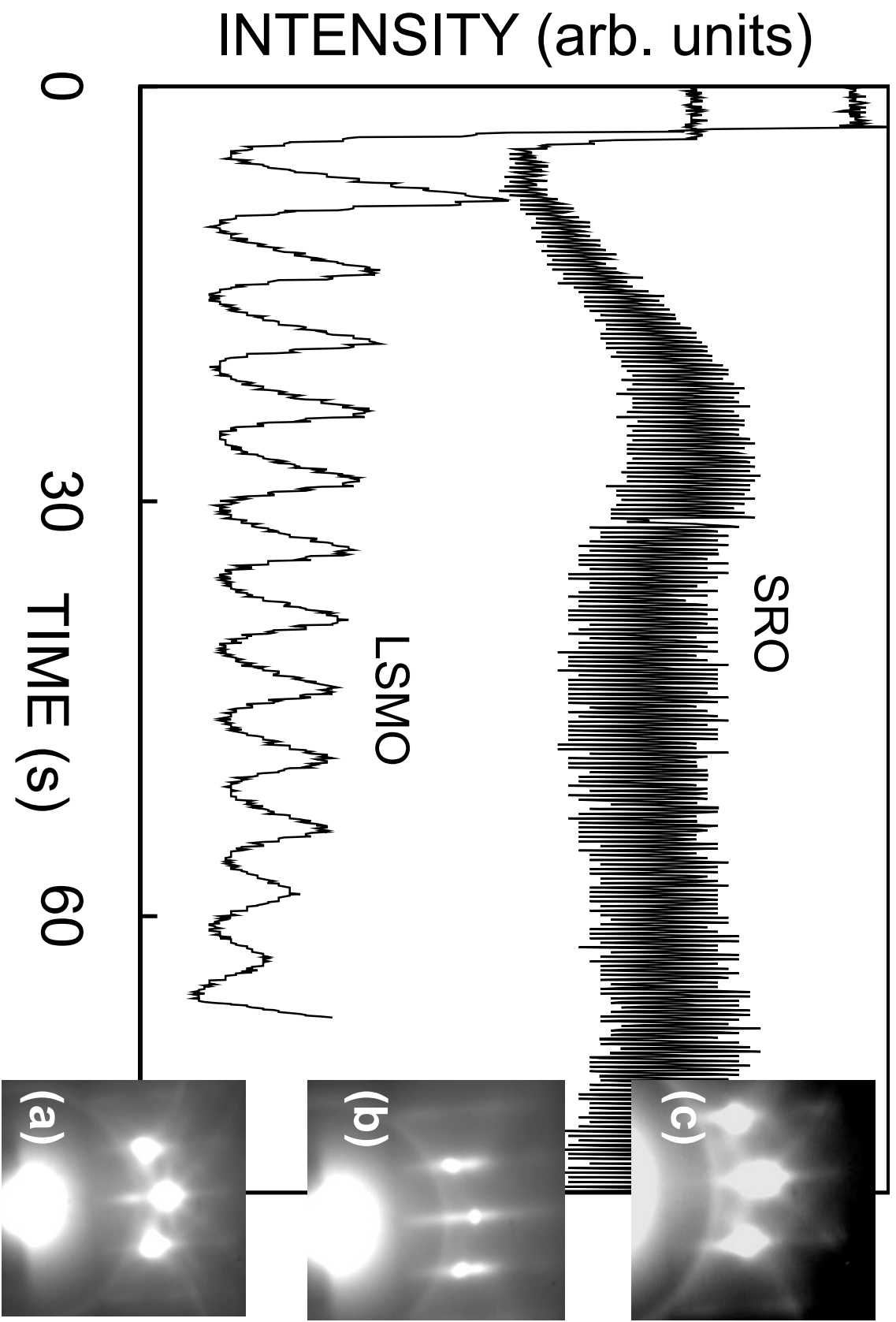

FIGURE 2

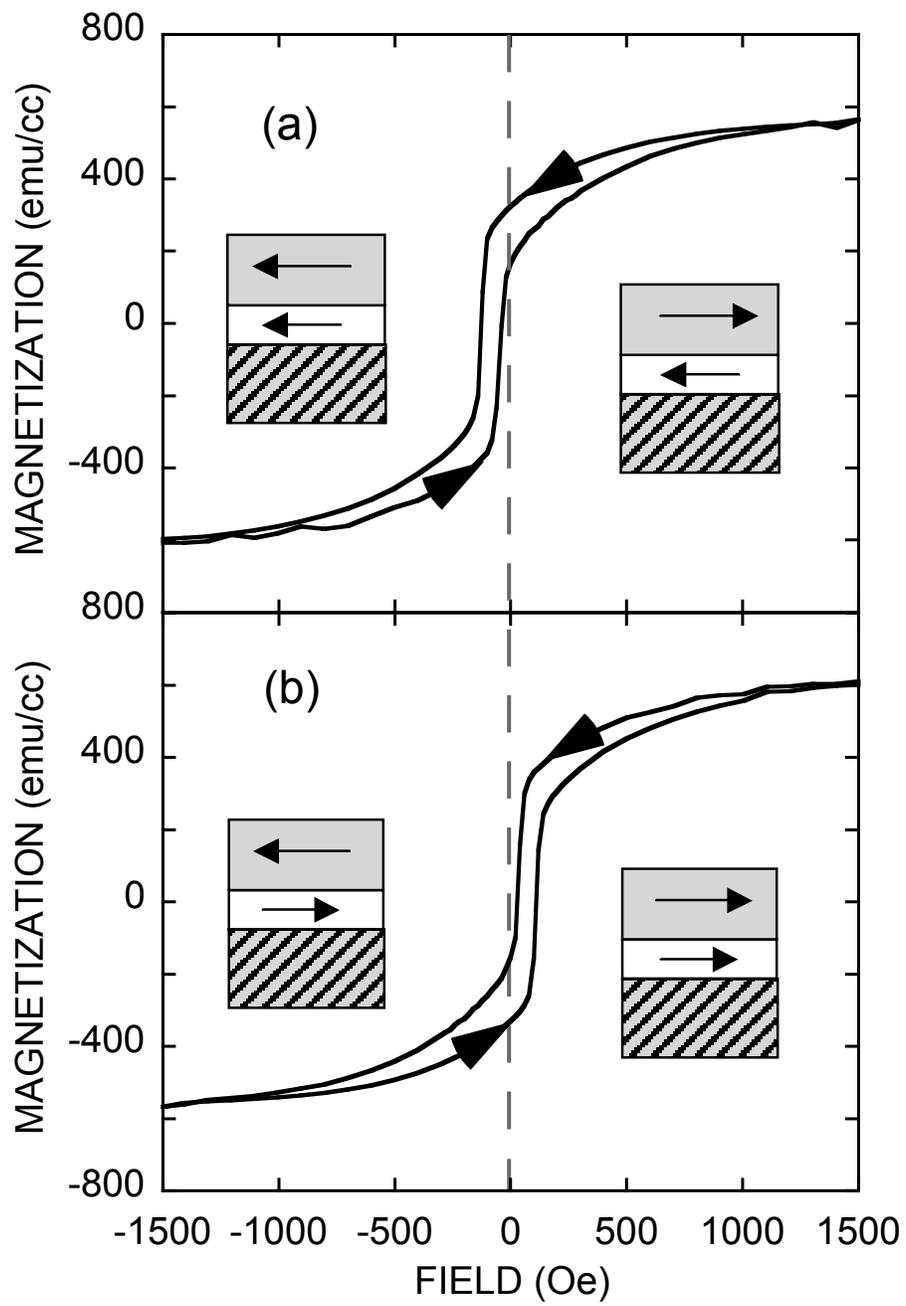

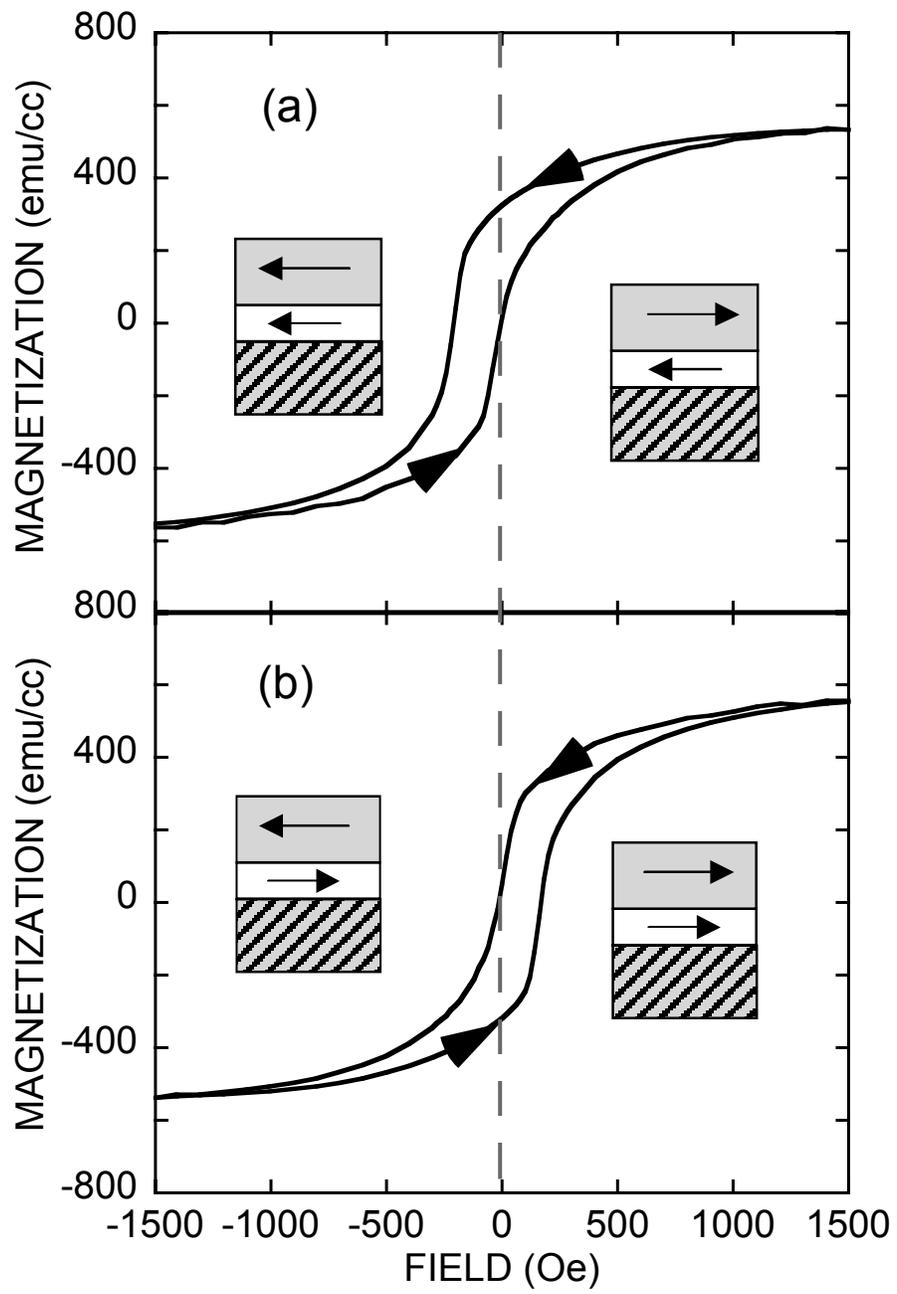

FIGURE 4

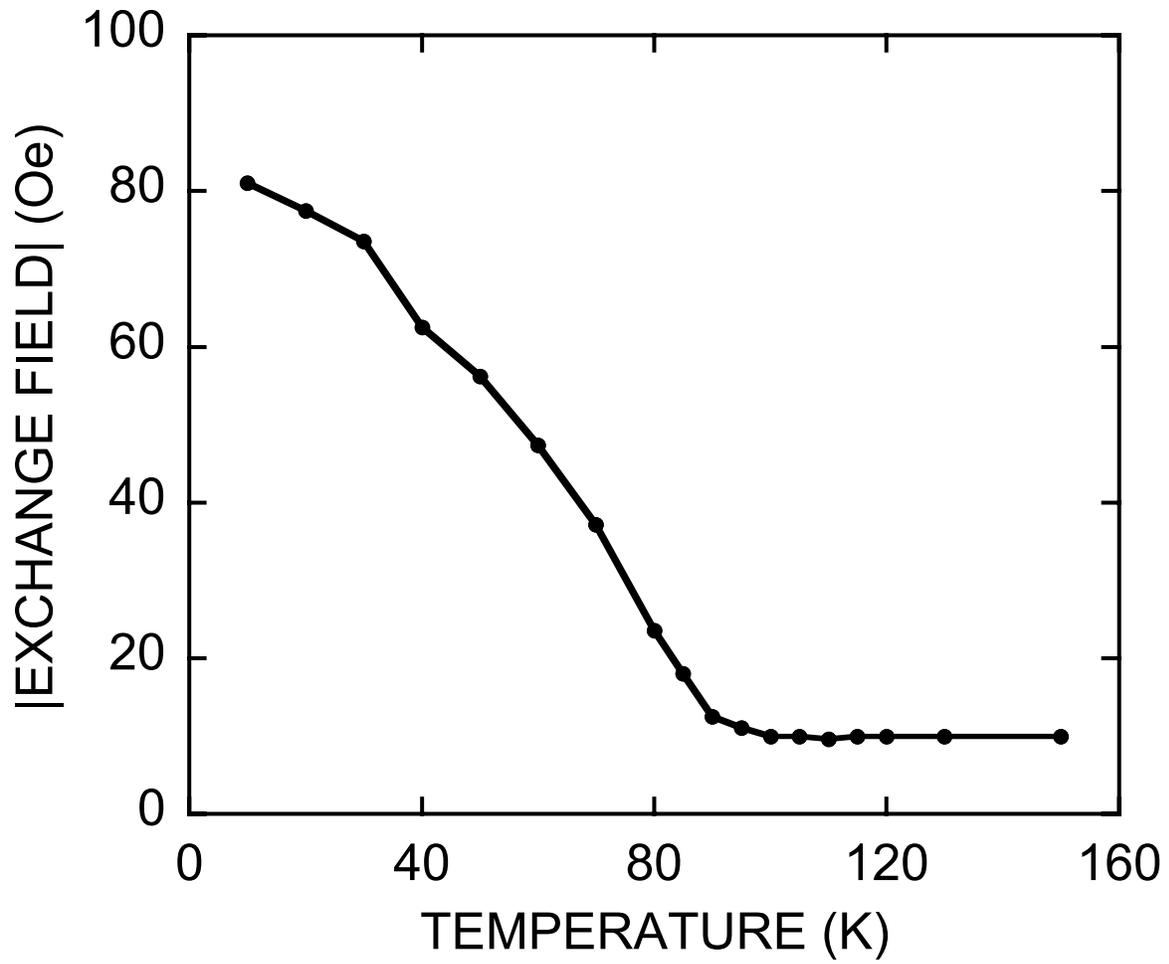